\documentclass{amsart}       
\usepackage{graphicx}
\usepackage{mathptmx}      
\usepackage{latexsym} 
\usepackage{amsmath} 
\usepackage{epsfig}
\usepackage{amssymb}
\usepackage{enumerate}
\usepackage{xcolor}

\usepackage{amsmath, amsfonts, amssymb} 
\usepackage{mathtools}
\usepackage{enumitem}
\usepackage[numbers, sort&compress]{natbib}
\usepackage[colorlinks]{hyperref}
\hypersetup{
	citecolor=blue,
   linkcolor=red,
}

\newtheorem{theorem}{Theorem}[section]
\newtheorem{lemma}[theorem]{Lemma}
\newtheorem{corollary}[theorem]{Corollary}

\newtheorem{proposition}[theorem]{Proposition}

\newcommand{\cB}{{\mathcal B}}

\newcommand{\cL}{{\mathcal L}}
\newcommand{\cM}{{\mathcal M}}
\newcommand{\cN}{{\mathcal N}}

\newcommand{\cT}{{\mathcal T}}

\title{Weakly displaying trees in temporal tree-child networks}

\author{Katharina T. Huber, Simone Linz, and Vincent Moulton}

	\address{School of Computing Sciences, University of East Anglia, Norwich, UK}
  \email{K.Huber@uea.ac.uk} 
	
\address{School of Computer Science, University of Auckland, Auckland, New Zealand}
 \email{s.linz@auckland.ac.nz}
	
	\address{School of Computing Sciences, University of East Anglia, Norwich, UK }
 \email{v.moulton@uea.ac.uk}



\date{\today}

\begin{document}

\begin{abstract}
	Recently there has been considerable interest 
	in the problem of finding a 
	phylogenetic network with a minimum 
	number of reticulation vertices which displays 
	a given set of phylogenetic trees, that is,
	a network with minimum hybrid number. Even so,
	for certain evolutionary scenarios insisting that a
	network displays the set of trees can be 
	an overly restrictive assumption. 
	In this paper, we consider the less restrictive 
	notion of displaying called weakly displaying and,
        in particular,
		a special case of this which we call rigidly displaying.  We
	characterize when two trees can be rigidly displayed 
	by a  temporal tree-child network in terms of fork-picking sequences,
	a concept that is closely related to that of cherry-picking sequences.
	We also show that, in case it exists, the rigid hybrid number for two phylogenetic trees
	is given by a minimum weight fork-picking sequence
	for the trees,  and that the rigid hybrid number can be quite 
	different from the related beaded- and temporal-hybrid numbers.
\end{abstract}

\maketitle

\section{Introduction}

Phylogenetic networks are a generalization of evolutionary trees.
They come in various forms, and are commonly used 
to represent the evolutionary history for a set $X$ of species in 
which events such as hybridization or recombination are suspected to 
have occurred \cite{huson}. For this paper, a phylogenetic network (on species set 
$X$) is  a connected directed acyclic graph, with a single root vertex and leaf-set $X$
in which every internal vertex has degree 3 except for
the root which has outdegree 2. We call the number of vertices in a 
phylogenetic network with indegree 2 
the network's {\em reticulation number}, so that
a {\em phylogenetic tree} is a network with reticulation number 0. 
We shall mainly focus on  {\em temporal tree-child} networks in which each non-leaf vertex
has a child whose indegree is 1, whose vertices 
	can be labelled with times that move strictly 
	forward on treelike parts of the network and so that vertices with
	indegree 2 have parents with the same time label (also
known as tree-child, time-consistent networks \cite{CLRV09}).

Any phylogenetic network on the set $X$ {\em displays} a set of 
	phylogenetic trees on $X$, 
	where a phylogenetic tree is {\em displayed} by
	a network if there is a subgraph of the network that is isomorphic to
	a subdivision of the tree \cite{vi2010}.
It is therefore natural to try to construct phylogenetic networks
by reversing this process, i.e. by trying to find a network
which displays a given set 
of trees. These trees are usually obtained from genomic data, 
by considering different genes (which leads to ``gene trees") 
or regions of the species' genomes.
For a given set of phylogenetic trees, this  also leads to the 
concept of the {\em (temporal) hybrid number}, which is the minimum 
reticulation number taken over all (temporal tree-child) networks that 
display each tree in the set \cite{BS,HLS13}. While the hybrid number 
exists for any set of phylogenetic trees, it is worth noting that the 
temporal hybrid number does not always exist, i.e. there 
are sets of trees that cannot simultaneously be embedded in a 
temporal tree-child network.

Several results have been presented in the 
literature concerning displaying phylogenetic trees and hybrid numbers, mainly
for pairs of trees.
These include structural information on how the hybrid number 
is related to the so-called {\em maximum acyclic agreement forest} for two phylogenetic trees  
\cite{B05}, characterizations for
when collections of trees are displayed by special types of networks \cite{HLS13,L19} 
and related algorithms/complexity results \cite{BS05,BS07,Dl17,HL13,H16}. 
However, all of these results rely on the fact that
the networks {\em display} the set of trees in question, a notion that  
may not appropriately model certain evolutionary
scenarios, such as incomplete lineage sorting \cite{L,degan}.

A possible solution to this problem is to 
relax the displaying condition. 
Roughly speaking, a phylogenetic tree is {\em weakly displayed} by a network \cite{huber16} if
it can be embedded in the network in such a way that
the tree follows along the directed paths in the network 
(see Section~\ref{sect:weakly} for the definition). 
In this paper we will consider the problem of deciding 
when a pair of phylogenetic trees is weakly
displayed by a temporal tree-child network 
under the assumption that there exist 
simultaneous embeddings of both trees that do not permit
more than three branches of the trees to 
come together at a reticulation vertex.
In this case we shall say that  the pair of trees 
is {\em rigidly displayed} by the network.
Note that related problems were recently 
considered in  \cite{L} (the Parental Tree Network Problem) 
and in \cite{IJ18} (The Beaded Tree Problem). In  
	the Parental Tree Network Problem
	the aim is to find a network with
	a minimum number of reticulation vertices that weakly displays all
	trees in a given set of phylogenetic  trees; 
	in the Beaded Tree Problem, however, networks with 
	parallel edges are permitted and a different 
	concept of displaying is used which can lead to different solutions
	(see  Section~\ref{sect:weaknumber} for more details).

We now summarize the rest of the paper, 
including statements of our main results. After presenting 
some definitions in Section~\ref{sect:prelims}, in Section~\ref{sect:weakly} 
we present the definition of weakly displaying, and 
we prove some basic facts 
concerning this concept and its relationship with displaying. 
In Section~\ref{sect:weaknumber}, 
we then consider the weak hybrid number of two trees. In
particular, we determine the weak hybrid number for a
specific pair of phylogenetic trees and show
that for this pair of trees we get a different number 
to the analogous hybrid number defined in \cite{IJ18}.
This example shows that the beaded trees introduced
in \cite{IJ18} can lead to a quite different solution
when aiming to find a network in which to embed the given trees.

In Sections~\ref{sect:rigidly} and \ref{sect:fork}, we introduce
the concepts of rigidly displaying and fork-operations, respectively
and prove some results on these concepts which we use later on. 
Then in Section~\ref{sect:forkpicking} we give a 
characterization for when a pair of phylogenetic trees can be rigidly displayed 
in terms of fork-picking sequences (Corollary~\ref{rigid-cherry}), a generalization of 
cherry-picking sequences  \cite{HLS13}. 
In Section~\ref{sect:characterize},
we go on to show that when the rigid hybrid number of two trees exists, it is 
equal to the weight of a minimum fork-picking sequence (Corollary \ref{rigidsize}).
These results can be regarded as
analogues of \cite[Theorem 1]{HLS13} and \cite[Theorem 2]{HLS13},  respectively.
In  Section~\ref{sect:relationship} we show that there is a
pair of phylogenetic trees on a set $X$ with $|X|$ arbitrarily large, so that  
the difference between the temporal and rigid hybrid numbers
for these two trees is at 
least $\frac{|X|}{4}-3$ (Theorem~\ref{big}).
We conclude with a discussion on possible future directions in 
Section~\ref{sect:discussion}.

\section{Preliminaries} \label{sect:prelims}

Let $G$ denote a directed, acyclic graph with a single root vertex,
i.e., a vertex
	with indegree 0.
	We let $V(G)$ denote the vertex set of $G$, $E(G)$
	the set of (directed) edges of $G$, and $\rho_G$ the unique root of $G$.
	A vertex in $G$ with indegree 1 and outdegree 0 is called a 
	{\em leaf}; an edge of $G$ incident with a leaf of $G$ a {\em pendant edge} of $G$.
	Furthermore, we denote the set of all leaves of $G$ by $L(G)$.

Suppose $v\in V(G)$. We say that a 
	vertex $u\in V(G)$ is {\em above}
	$v$ if there exists a directed path from the root of $G$ to $v$ that
	contains $u$ (note that $u$ could equal $v$).
	If $u$ is above $v$, then we also write $u\preceq_G v$ or simply
	$u\preceq v$ if $G$ is clear from the context.
	Furthermore, we also say that $v$ is {\em below} $u$.
	We call any vertex above $v$ an {\em ancestor} of $v$
	and any vertex below $v$ a {\em descendant} of $v$. Finally, we
	say that two distinct edges $e$ and $e'$ of $G$ are {\em comparable}
	if $head(e)$ is above  $tail(e')$ 
	or $head(e')$ is above  $tail(e)$. Otherwise we say
	that $e$ and $e'$ are {\em incomparable}.

Let $X$ be a finite set of size at least 2. 
Following e.g. \cite[p.1764]{huber16}
a rooted, directed acyclic graph  $\cN$ is called a
{\em phylogenetic network (on $X$)} if the leaf-set of $\cN$ is $X$,
the root $\rho_{\cN}$ of $\cN$ has outdegree
	two and any non-root, non-leaf vertex $v\in V(\cN)$ 
either has indegree one and outdegree two or zero
(in which case $v$ is called a {\em tree vertex})
or $v$ has indegree two and its outdegree is one
(in which case $v$ is called a {\em reticulation vertex}).
The set of reticulation vertices
of $\cN$ is denoted by $Ret(\cN)$. We put $h(\cN) = |Ret(\cN)|$.
Unless stated otherwise, phylogenetic networks do not
contain parallel edges. Moreover, we call a 
	directed path in a phylogenetic
	network $\cN$ of length one or more
	in which every vertex, except possibly the first vertex,
	is a tree vertex a {\em  tree-path} in $\cN$.

A {\em phylogenetic tree (on $X$)} is a phylogenetic network on $X$
that does not have any reticulation vertices. We say that
two phylogenetic trees $\cT$ and $\cT'$ on $X$ are {\em isomorphic},
denoted by $\cT\cong \cT'$,
if there exists a bijection $V(\cT)\to V(\cT')$
that induces a graph isomorphism between $\cT$ and $\cT'$
that is the identity on $X$. If $\cT$ is a phylogenetic tree on $X$, and 
	$Y\subseteq X$, then the {\em last common ancestor of $Y$},
	denote by $lca_{\cT}(Y)$, 
	is the unique vertex $v$ of $\cT$ that is 
	an ancestor of every element in $Y$ and 
	there is no vertex $w$
	such that $w$ is a descendant of $v$ and $w$ 
	is an ancestor of every element in $Y$.
	For any $2\leq l\leq |X|$ elements $x_1,\ldots x_l\in X$,
	we sometimes also write $lca_{\cT}(x_1,\ldots, x_l)$ rather than
	$lca_{\cT}(\{x_1,\ldots, x_l\})$.
	In addition, we denote by $\cT(Y)$ the minimal
	subtree of $\cT$
	spanned by all leaves in $Y$ and by $\cT|_Y$ the {\em restriction} of $\cT$
	to $Y$, that is, the phylogenetic tree on $Y$
	obtained from $\cT(Y)$ by suppressing all resulting vertices of both indegree
	and outdegree one. Note that the root of $\cT(Y)$ is the 
	last common ancestor of all elements in $Y$.

Suppose $\cN$ is a phylogenetic network on $X$.
Following \cite{S16},  we say that
$\cN$ is {\em temporal} \cite{M04} if there exists a map $t : V(\cN)\to \mathbb R^{\geq 0}$
such that, for all 
$(p,q)\in E(\cN)$, we have $t(p) = t(q)$ whenever
$q$ is a reticulation vertex and $t(p) < t(q)$, otherwise.
In that case, we call $t$ a {\em temporal labelling} of $\cN$.
Unless of relevance to the discussion, we always omit the
temporal labelling when depicting a temporal network.
We say that $\cN$ is {\em tree-child} \cite{CRV09} if, for each non-leaf vertex
$v \in V(\cN)$  at least one of the children of $v$ is a tree vertex.
Note that a tree-child network was called a
phylogenetic network in \cite[p.1883]{HLS13}. Also note
that a temporal tree-child network (in our sense) has also been called a
(binary) {\em time-consistent tree-child network}, or {\em TCTC-network}
in \cite{CLRV09}. Finally, we say that $\cN$ is {\em normal} 
if in addition to being tree-child it does not contain a
{\em shortcut}, that is, if there is a directed
path from a vertex $u\in V(\cN)$ to a vertex $v\in V(\cN)$
with at least two edges, then there is no
directed edge $(u,v)$ \cite{MS15}.  Note that any temporal
tree-child network is normal \cite[Proposition 10.12]{S16}

\section{Weakly displaying two trees in a network}\label{sect:weakly}

In this section, we derive some basic properties 
for the notion of weak displaying  which will be useful later.
In  the following, assume that $\cT$ is a phylogenetic tree on $X$ and
that $\cN$ is a phylogenetic network on $X$.

We call a map $ \psi:V(\cT)\to V(\cN)$ that is
the identity on $X$ a {\em display map}  for $\cT$
in $\cN$ if the following additional properties hold
\begin{enumerate}
	\item[(i)] for all $v\in V(\cT)$, $\psi(v)$ is a tree vertex or the
	root of $\cN$,
	\item[(ii)]  for every edge $e$ of $\cT$ there exists a directed
	path $\psi[e]$ having at least one edge
	from $\psi(tail(e))$ to $\psi(head(e))$
	in $\cN$, and
	\item[(iii)] for any two distinct edges $e$ and $e'$ of $\cT$
	that share the same tail
	the first edge of
	$\psi[e]$ is not the first edge of
	$\psi[e']$.
\end{enumerate}
Following \cite{huber16}, we say that $\cT$ is
{\em weakly displayed} by $\cN$ if there exists a display map for $\cT$
in $\cN$. To reduce notation, we will sometimes
	not explicitly refer to the display map. 
Note that if $\cT$ is weakly displayed by $\cN$,
then there could be more than one display map for $\cT$
in $\cN$.  In addition, note that if $\cN$ displays $\cT$ then $\cN$ also 
weakly displays $\cT$ (but not necessarily conversely).

The notion of weakly displayed was 
introduced in \cite{huber16} in terms of a construction that
allows the ``unfolding'' of a phylogenetic network on $X$ into a so-called
``multi-labelled tree on $X$'' \cite{HM06}. Such trees
are similar to phylogenetic trees in that they have no vertices
with in- and outdegree one and the root has indegree zero. However
the requirement that the leaf-set is $X$ is relaxed to the requirement
that an element of $X$ can ``label'' more than one leaf (which is not
allowed in the case of  phylogenetic trees).


Note that although closely related,  display maps are not
{\em weak embeddings} sensu \cite{IJ18}. Stated within our
framework, such embeddings are maps
$\psi:V(\cM)\to V(\cN)$ from a multi-labelled tree $\cM$ into a
phylogenetic network $\cN$ such that (a) all leaves of $\cM$ that share
the same ``label'' $x\in X$  are mapped to the leaf $x$ of $\cN$,
(b) every edge $e$ of $\cM$ is mapped to either a vertex of $\cN$ or
a directed path from $\psi(tail(e))$ to $\psi(head(e))$, and (c) for every
non-leaf vertex $v$ of $\cM$ with outgoing edges $a_1$ and $a_2$, 
the directed paths associated to $a_1$ and $a_2$ which each
has at least one edge)
start with different outgoing edges of $\psi(v)$. Since a phylogenetic
	tree is clearly also a multi-labelled tree, it follows that the map
	$\psi: V(\cT)\to V(\cN)$ that is the identity on $X$ and maps all
	non-leaf vertices of $\cT$ to the root of $\cN$ is a weak embedding
	of $\cT$ into $\cN$ as Property~(c) vacuously applies. However
	$\psi$ is not a display map for $\cT$ in $\cN$ as Property~(ii) does not hold.
	Even so, a display map is always a weak embedding.

\begin{figure}[t]
	\center
	\scalebox{0.9}{\includegraphics{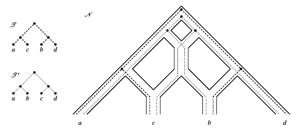}}
	\caption{Two phylogenetic trees $\cT$ and $\cT'$ on
		$X=\{a,b,c,d\}$ that are weakly
		displayed by the network $\cN$ on $X$ for which
		$\gamma(w) \le 2$ holds for all $w \in V(\cN)$. However,
		$\cT$ and $\cT'$ are {\em not} both
		displayed by $\cN$ (the tree $\cT'$ is not displayed).}
	\label{fig:nonexample}
\end{figure}

Now, suppose that $\psi:V(\cT)\to V(\cN)$
	is a display map for $\cT$ in $\cN$. For 
	any edge $e=(u,v)\in E(\cT)$, we denote by $\psi[e]-\psi(u)$ the set
	of all vertices in $V(\cN)$ that lie on 
	the path $\psi[e]$ except for $\psi(u)$.
	Let $w \in V(\cN)$, and let $e=(u,v)$ be an edge of $\cT$.
	If $\psi(v)=w$, we
	say that  the path {\em $\psi[e]$  ends at $w$} and if
	$w \in \psi[e]-\psi(u)$, but $\psi(v)\neq w$
	we say that {\em $\psi[e]$  passes through $w$.}
	In addition, we define the number
	$$
	\gamma_{\psi}(w)=|\{e=(u,v) \in E(\cT) \,:\,
	w \in \psi[e]-\psi(u) \}|,
	$$
	i.e., $\gamma_{\psi}(w)$ counts the number of edges in
	$\cT$ such that their
	image under $\psi$ either ends or passes through $w$
	so that in particular $\gamma_{\psi}(\rho_{\cN})=0$. Finally, if $\cT'$ is 
	a further phylogenetic tree on $X$ that is also
	weakly displayed by $\cN$ via a map $\psi'$, then we put 
	$$
	\gamma(w)=
	\gamma_{\psi,\psi'}(w)=
	\gamma_{\psi}(w)+\gamma_{\psi'}(w)
	$$ 
	\noindent (see e.g. Figure~\ref{fig:nonexample}). To reduce
	notation we sometimes drop the subscript in $\gamma_{\psi,\psi'}$ 
	as indicated when no confusion can arise 
	about which maps are being used to weakly display $\cT$ and $\cT'$.

We now prove two lemmas about these concepts which
	will be useful later. The first concerns temporal tree-child
	networks.

\begin{lemma}\label{display} Let $\cN$ be a temporal tree-child
	network on
	$X$ that weakly displays a phylogenetic tree $\cT$ on $X$ via a display map $\psi$.
	Then $\cN$ displays $\cT$ if and only if for all $w \in V(\cN)$, we have $\gamma_{\psi}(w)\le 1$.
	In addition, $\psi_{\cT}(\rho_{\cT}) = \rho_{\cN}$ so that, in particular, if $\cN$
	displays $\cT$ then $\psi_{\cT}(\rho_{\cT}) = \rho_{\cN}$.	
\end{lemma}
\noindent {\em Proof:} Consider the first statement. Suppose $w \in V(\cN)$, then
by the definition of displaying, it is straight-forward to see that, for all
$w \in V(\cN)$, we have  $\gamma_{\psi}(w)\le 1$. Conversely, 
	first note that if $w \in Ret(\cN)$ then, by assumption,
there exists at most one
edge $e\in E(\cT)$ such that $\psi_\cT[e]$ either passes through
or ends in $w$. As $w$ is a reticulation vertex, $\psi_\cT[e]$ cannot
end at $w$. Since $\cN$ has no shortcuts, 
by deleting each edge $e$ of $\cN$
that is directed into a reticulation vertex $w'$ of $\cN$ and for
which there exists no $e'\in E(\cT)$ such that $e$ is an edge on
$\psi_\cT[e']$  and
$\psi_\cT[e']$  passes through $w'$, we obtain a subgraph of $\cN$
that is isomorphic to a subdivision of $\cT$. Thus, $\cN$ displays $\cT$.

To see that the last statement in the lemma holds, 
	assume for contradiction that
$\psi_{\cT}(\rho_{\cT})$ is not the root $ \rho_{\cN}$ of $\cN$.
Since $\cN$ is temporal tree-child  and, therefore,
normal \cite[Proposition 10.12]{S16}, the two children $u$ and $v$
of $\rho_{\cN}$ must be distinct and tree vertices. 
Moreover, there must exist a tree-path $p_x$ in $\cN$ from $u$ to 
some leaf $x$ and a tree-path $p_y$ from $v$ to some leaf $y$.
Note that since $p_x$ and $p_y$ cannot intersect, we must have
$x\not = y$.
Since $\psi_{\cT}$ is a display map
for $\cT$ in $\cN$ and $x$ and $y$ are also
leaves of $\cT$ it follows that
$lca_{\cT}(x,y)$ is mapped to an ancestor of
$x$ and $y$ in $\cN$ under $\psi_{\cT}$.
Extending the paths $p_x$ and $p_y$ to tree-paths
starting at $\rho_{\cN}$ implies that that ancestor must be
$\rho_{\cN}$. Thus, $\psi_{\cT}(\rho_{\cT}) = \rho_{\cN}$.
\qed

To state the second lemma  we require some further definitions.
We call a subgraph
$\cN'$ of $\cN$ a {\em pendant subnetwork} of $\cN$ if there exists a
tree vertex $v$ in $\cN$ such that when deleting the incoming
edge of $v$ the network
decomposes into two connected components such that
the component that contains $v$ in its
vertex set is a phylogenetic network $\cN'$ on
$L(\cN')\subseteq L(\cN)$. A {\em pendant subtree} of $\cN$
is a pendant subnetwork of $\cN$ that is a phylogenetic tree.
Note that a pendant subnetwork and therefore also a pendant
subtree must have at least two leaves.

\begin{lemma}\label{helper}
	Suppose that $\cN$ is a phylogenetic network on $X$ that
	weakly displays two distinct phylogenetic trees $\cT$ and $\cT'$ on $X$ via
		display maps $\psi$ and $\psi'$, respectively. 
	\begin{itemize}
		\item[(i)] If $v$ is a tree vertex in $\cN$ and
		$\gamma_{\psi,\psi'}(v) \ge 3$, then there is a vertex $w \in Ret(\cN)$
		which is an ancestor of $v$ in $\cN$.
		\item[(ii)] If $v$ is tree vertex in $\cN$ with $\gamma_{\psi,\psi'}(v)=2$
		and $v$ has a child that is the root of a pendant subtree,
		say $\cT^*$, of $\cN$
		then $\cT^*$ is a pendant subtree of
		both $\cT$ and $\cT'$. 
	\end{itemize}\end{lemma}
	\noindent {\em Proof:}
	(i) Suppose that $v$ is a tree vertex of  $\cN$ and
	$\gamma(v) \ge 3$. Then, without loss of generality, we may assume that
	$\cT$ is such that
	$\gamma_{\psi}(v)\ge 2$.
	Hence, there are two distinct edges $e$ and $e'$ in $\cT$ with
	tails $u$ and $u'$ respectively, such that
	$v \in \psi[e]-\psi(u)$
	and $v \in \psi[e']-\psi(u')$.
	Note that $u\not=u'$ may or may not hold.
	In either case, the definition
	of a display map combined with the fact that $\cN$ does
	not contain parallel edges implies that the heads of the outgoing edges
	of $\psi(u)$ and  $\psi(u')$, respectively, must be distinct.
	Thus, $\psi[e']\not=\psi[e]$.
	
	We claim that $e$ and $e'$ are incomparable in $\cT$. Indeed, assume for
	contradiction that $e$ and $e'$ are comparable. 
	Without loss of generality we may assume that
	$head(e)$ is above $tail(e')$ in $\cT$. Then the 
	directed path $P$ in $\cT$ starting at $u$ and ending at the head of $e'$ 
	is mapped by $\psi$ to a directed path in $\cN$
	with edge set $\bigcup_{e\in E(P)}E(\psi[e])$.
	Since $v \in \psi[e]-\psi(u)$ it follows that 
	$v \not\in \psi[e']-\psi(u')$, a contradiction which
	yields the claim.
	
	In particular, since $\cT$ is weakly displayed by $\cN$ there must be a 
	vertex $q$ in $\cT$ with $q \preceq_{\cT} u$ and $q \preceq_{\cT} u'$ such that
	(a) the two directed paths  in $\cT$ from $q$ up to and including the heads
	of $e$ and $e'$, respectively, are mapped by $\psi$ to 
	two directed paths $p_e$ and $p_{e'}$ in $\cN$, (b) the first
	edge on $p_e$ is different from the first edge on $p_{e'}$,
	and (c), $v$ is a vertex on both $p_e$ and $p_{e'}$.
	Hence there must be some vertex in $Ret(\cN)$ 
	which lies on $p_e$ and $p_{e'}$ and which is an
	ancestor of $v$.
	
	\noindent (ii) Note that every leaf $x$ in $\cT^*$ must be contained in the
	image under $\psi$ of some directed
	path in $\cT$ from the root of $\cT$ to $x$,  and similarly for $\cT'$.  
	Since, by assumption, $\gamma(v)=2$, there can be at
	most one such path in $\cT$ and $\cT'$, respectively which has this
	property for every leaf in $\cT^*$. Hence $\cT^*$ must be a pendant
	subtree of both $\cT$ and $\cT'$.
	\qed

	\section{The weak hybrid number}\label{sect:weaknumber}
	
	Given two phylogenetic trees $\cT$ and $\cT'$ on $X$, we define the {\it weak 
		hybrid number} $h_{wd}(\cT,\cT')$ of $\cT$ and $\cT'$ as
	\begin{eqnarray*}
		&\phantom{f}& h_{wd}(\cT,\cT')\\
		&=&
		\min\{h(\cN): \cN \text{ is a phylogenetic network
			that weakly displays }\cT\text{ and }\cT'\}.
	\end{eqnarray*}
	Note that for any two phylogenetic trees $\cT$ and $\cT'$
		there always exists a phylogenetic network that displays
		$\cT$ and $\cT'$ and so $h_{wd}(\cT,\cT')$ is well-defined.
		In addition, the  weak hybrid number has been implicitly considered in 
		The Parental Tree Network Problem \cite[Definition 5]{L}.
	In this section, we give an example which shows that
	the weak hybrid number is different from the related
	beaded hybrid number \cite{IJ18}, whose definition 
		we next recall. 
	
	A {\it beaded tree} $\cB$ on $X$ is a phylogenetic network
		on $X$ in which parallel edges are allowed, and in which
		each reticulation $v$ has a unique parent $u$ such that
		there are two parallel edges from $u$ to $v$ \cite[Definition 7]{IJ18}. 
	Now for two phylogenetic
	trees $\cT$ and $\cT'$ on $X$, we define the {\it beaded hybrid number}
	$h_b(\cT,\cT')$ for $\cT$ and $\cT'$ to be
	\begin{eqnarray*}
		h_b(\cT,\cT')&=&\min\{h(\cB): \cB \text{ is a beaded tree such that
			there exist weak }\\
		&& \text{ embeddings of 
		} \cT\text{ and }\cT' \text{ into } \cB\}.
	\end{eqnarray*}
	Note that  
		\cite[Lemma 9]{IJ18} implies that any phylogenetic
		network $\cN$ on $X$ that weakly displays two phylogenetic trees
		$\cT$ and $\cT'$ on $X$ can be transformed into a beaded tree
		$\cB$ on $X$ such that there exist weak embeddings of $\cT$ and $\cT'$
		into $\cB$ for which $|Ret(\cN)|=|Ret(\cB)|$ (so in particular 
		$h_b(\cT,\cT')$ exists for any pair of trees $\cT,\cT'$).
		Hence, $h_b(\cT,\cT') \le h_{wd}(\cT,\cT')$.
	
\begin{figure}[t]
		\center
		\scalebox{0.99}{\includegraphics{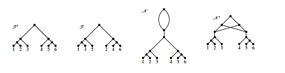}}
		\caption{Two phylogenetic trees $\cT$
			and $\cT'$ on $X=\{1,\ldots, 6\}$
			whose beaded hybrid number $h_b(\cT,\cT')$ is 1 ($\cN$ 
				is a beaded tree into which $\cT$ and $\cT'$ can be weakly embedded), 
			and whose
			weak hybrid number $h_{wd}(\cT,\cT')$ is  2
			(the network $\cN'$ 
				weakly displays both trees).}
		\label{fig:beaded}
	\end{figure}

	We now use Lemma~\ref{helper} to show in Proposition~\ref{nastyexample}
	that there exist phylogenetic trees $\cT$ and $\cT'$  such that
	$h_{wd}(\cT,\cT')\not=h_b(\cT,\cT')$. First,
	consider the two phylogenetic trees $\cT$ and $\cT'$ depicted
	in Figure~\ref{fig:beaded}. Note that $h_b(\cT,\cT')=1$ 
		since $\cT$ and $\cT'$ are not isomorphic, and 
		there exist weak embeddings of $\cT$ and $\cT'$ into
		the pictured beaded tree $\cN$, respectively. Hence, $h_{wd}(\cT,\cT')\geq 1$.
	We now show that  $h_{wd}(\cT,\cT')=2$. 
	
	To this end, we call
	two leaves $x$ and $y$ of a phylogenetic tree $\cT$ with $x\not=y$
	a {\em cherry} of $\cT$, denoted by $\{x,y\}$, if $x$ and $y$ share
	a parent. 
	
	\begin{proposition}\label{nastyexample}
		Let  $\cT$ and $\cT'$ denote the two phylogenetic trees on
		$X=\{1,\ldots, 6\}$
		pictured in Figure~\ref{fig:beaded}. Then $h_{wd}(\cT,\cT')= 2$.
	\end{proposition}
	\noindent {\em Proof:} 
	As $\cT$ and $\cT'$ are not isomorphic we  have
	$h_{wd}(\cT,\cT') \ge 1$. Moreover, as the phylogenetic network
	$\cN'$ pictured  in Figure~\ref{fig:beaded} is also on $X$ and weakly displays 
	$\cT$ and $\cT'$ we  have $h_{wd}(\cT,\cT') \le 2$. We now show that 
	$h_{wd}(\cT,\cT') \neq 1$, from which the proposition follows.
	
	Suppose to the contrary that $h_{wd}(\cT,\cT') =1$.
	Then there exists a phylogenetic network $\cN^*$ that weakly 
	displays $\cT$ and $\cT'$ such that $h(\cN^*)=1$.
	Let $v$ be the unique vertex in $Ret(\cN^*)$.
	Let $u\in V(\cN^*)-\{\rho_{\cN^*}\}$ be a parent of $v$.
	Note that since $\cN^*$
	has no parallel edges, $u$ must exist. Also note that $u$
	must be a tree vertex of $\cN^*$ as
	$v$ is the sole reticulation vertex of $\cN^*$. Finally, note that
	the other child of $u$ cannot be $v$ as
	$\cN^*$ does not contain parallel edges.

        Denoting that child
	by $x$ we next claim that $x$ must be a leaf of $\cN^*$. 
        Assume for contradiction that $x$ is not a leaf. Let $\cT^*$ the
        subtree of $\cN^*$ rooted at $x$. Let $x'$ be a
        leaf of $\cT^*$
	and, thus, of $\cN^*$. Then since $\cN^*$
	weakly displays $\cT$ via a map $\psi$ say, and
          $v$ is the sole reticulation vertex of $\cN^*$
		we obtain $\gamma_{\psi}(u)\geq 1$. Similarly, as $\cN^*$
		weakly displays $\cT$ via a map $\psi'$ say, 
		$\gamma_{\psi'}(u)\geq 1$ must hold. Thus,
                $\gamma_{\psi,\psi'}(u)\geq 2$. Since
		Lemma~\ref{helper}(i) implies that $\gamma_{\psi,\psi'}(u)\leq 2$
		as  $v$ is the sole reticulation vertex of $\cN^*$, it follows
		that $\gamma_{\psi,\psi'}(u)=2$.  Hence, by
        Lemma~\ref{helper}(ii),
	$\cT^*$ is also a pendant subtree of $\cT$ and of $\cT'$;
	a contradiction as $\cT$ and $\cT'$ are the two trees depicted in
	Figure~\ref{fig:beaded}. Thus, $x$ is a leaf of $\cN^*$, as claimed.

	
	Since every element in $X$ is contained in
	a cherry of either $\cT$ or $\cT'$, we may choose some $y \in X-\{x\}$
	such that $\{x,y\}$ is a cherry in either $\cT$ or $\cT'$. Without
	loss of generality, assume that $\cT$ is that tree. Since
	the only two cherries of $\cT$ are $\{1,3\}$ and $\{4,6\}$
	we may assume without loss of generality that $\{x,y\}=\{1,3\}$.
	Let $m$ denote the parent of $x$ and $y$ in $\cT$.
	
	Let $w\in V(\cN^*)$ be the parent of $u$ which must exist as
	$u\not=\rho_{\cN^*}$. Then  $w\not=\rho_{\cN^*}$ as otherwise
	the fact that $\{x,y\}$ is a cherry of $\cT$ but not of $\cT'$ implies that
	$y$ is below $v$. But then $\cT$ is not
	weakly displayed by $\cN^*$ because $(\rho_{\cN^*},u)$ is an
	edge of $\cN^*$ and $(\rho_{\cT}, m)$ is not an edge in
	$\cT$; a contradiction.
	
	We next claim that $(w,v)$ cannot
	be an edge in $\cN^*$. To see this, assume for contradiction
        that  $(w,v)$
	is an edge in $\cN^*$. Then since $\cT$ is weakly displayed by $\cN^*$
	and $x$ is contained in a cherry of $\cT$ but not of $\cT'$
	it follows that $y$ must be a leaf of $\cN$ below $v$. If there
	existed another leaf of $\cN^*$ below $v$ then that leaf would
	have to be ``5''. Since $\{5,6\}$ is a cherry of $\cT'$
	and $\cT'$  is weakly displayed by $\cN^*$ it follows that
	that cherry must also be below $v$; a contradiction as 
	$\cT$ is the phylogenetic trees depicted in Figure~\ref{fig:beaded}.
	Thus, $y$ is in fact the sole leaf of $\cN^*$ below $v$. But then
	$\{x,y\}$ must also be a cherry of $\cT'$; a contradiction
	since  $\cT$ and $\cT'$ are the
	phylogenetic trees depicted in Figure~\ref{fig:beaded}.
	Thus,  $(w,v)$ cannot be an edge in $\cN^*$, as claimed.
	Hence,  the other child of $w$, call it $z$, must either be a leaf of
	$\cN^*$  or is the root of a pendant subtree of $\cN^*$.

        Note first that  arguments similar to the case of $x$
          imply that  $z$ must be a leaf of $\cN^*$.
	Let $p\in V(\cN^*)$ denote the parent of $w$.
	We next distinguish between the cases that
	$z=y$ and that $z\not=y$.
	
	If $z=y$ then $p\not=\rho_{\cN^*}$. To see this, assume for contradiction
	that $p=\rho_{\cN^*}$. Then $(\rho_{\cN^*},w)$ is an
	edge in $\cN^*$. Since $\{x,y\}$ is a cherry of $\cT$
	and the parent $m$ of $x$ and $y$ is not adjacent with $\rho_{\cT}$
	it follows that $\cT$ is not weakly displayed by $\cN^*$; a contradiction.
	Thus, $p\not=\rho_{\cN}$, as required.
	
	We next claim that $(p,v)$ also cannot be an edge of
	$\cN^*$. Assume for contradiction that $(p,v)$ is an edge of
	$\cN^*$. Then since $\{x,y\}$ is a cherry of $\cT$
	and $\cT$ is weakly displayed by $\cN^*$, similar arguments
	as before imply that ``5'' must be the sole 
	leaf of $\cN^*$ below $v$ and that the unique 
	directed path from $\rho_{\cN^*}$ to leaf ``2'' does not
	cross $w$. Since $\cT'$ is one of the two phylogenetic trees
	depicted in Figure~\ref{fig:beaded} it follows that $\cT'$ is not weakly
	displayed by $\cN^*$; a contradiction. Thus,
	$(p,v)$ cannot be an edge of $\cN^*$ either.
	
	Let $q$ denote the other child of $p$.
	Then similar arguments as in the case of $z$
	imply that $q$ must also be a leaf of $\cN^*$.
	Hence,  $(q,(x,y))$ is a pendant subtree of $\cT$.
	Since $p$ is a tree vertex of $\cN^*$, Lemma~\ref{helper}(ii)
	implies that $(q,(x,y))$ is a pendant subtree of $\cT$ and of $\cT'$;
	a contradiction in view of Figure~\ref{fig:beaded}.
	Hence, $h_{wd}(\cT,\cT') \neq 1$ in case $z=y$.
	
	Assume for the remainder that $z\neq y$. Then $y$  is a descendant 
	of $v$ in $\cN^*$. Since $z$ is a leaf of $\cN$, it follows that
	$(z,(x,y))$ is a pendant subtree of $\cT$. Since $\{x,y\}= \{1,3\}$
	we must have $z=5$. But then $x=6$ as $\{5,6\}$ is a cherry
	of $\cT'$ and $\cT'$ is weakly displayed by $\cN^*$; a final contradiction.
	Hence, $h_{wd}(\cT,\cT') \neq 1$ and, so, the proposition follows.
	\qed
	
	This example is important as it indicates that the beaded hybrid number could potentially
	underestimate the number of reticulations required to weakly display two phylogenetic trees in a network.
	It would be interesting to understand how large the difference  between 
		$h_b(\cT,\cT')$ and $h_{wd}(\cT,\cT')$  could be in general.
	
	\section{Rigidly displaying}\label{sect:rigidly}
	
	We now introduce and present some basic properties of the notion of
		rigidly displaying. We begin with a
		lemma which will help to motivate our definition. 
		In Figure~\ref{fig:nonexample} we present an example where two
		phylogenetic trees $\cT$ and $\cT'$ are weakly 
		displayed by the depicted phylogenetic network $\cN$,
		$\gamma(w) \le 2$ for all $w \in V(\cN)$ but
		$\cT$ and $\cT'$ are {\em not} both 
		displayed by $\cN$. So, in general, it 
		does not suffice to insist that $\gamma(w) \le 2$ for all $w \in V(\cN)$ 
		for  two phylogenetic  trees to be displayed by a phylogenetic network.
		However, if we insist that the network is 
		temporal tree-child, we now show that this condition actually suffices.
	
	\begin{lemma}\label{display-gamma}
		Suppose that $\cN$ is a temporal tree-child network on $X$  and
		that $\cT$ and $\cT'$ are two phylogenetic trees on $X$ that are weakly
		displayed by $\cN$ via display maps $\psi$ and $\psi'$, respectively.
		Then the following statements are equivalent. 
		\begin{itemize}
			\item[(i)] $\cN$ displays $\cT$ and $\cT'$.
			\item[(ii)] $\gamma_{\psi,\psi'}(v)=2$ for all $v \in V(\cN)-\{\rho_{\cN}\}$.
			\item[(iii)] $\gamma_{\psi,\psi'}(v)=2$ for all $v \in Ret(\cN)$.
		\end{itemize}
	\end{lemma}
	\noindent {\em Proof:}
	(i) $\Rightarrow$ (ii)
	We show first that $\gamma(v)\geq 2$ must hold
	for all  $v \in V(\cN)-\{\rho_{\cN}\}$. Assume for contradiction that
	there exists some vertex $v\in V(\cN)$ such that $\gamma(v)\leq 1$.
	Then one of $\gamma_{\psi}(v)=0$ or $\gamma_{\psi'}(v)=0$
	must hold. Without loss of generality we may assume that
	$\gamma_{\psi}(v)=0$. Then there exists no edge $e\in E(\cT)$
	such that $\psi[e]$ either passes through $v$ or ends in $v$.
	But then there cannot exist a leaf $x$ of $\cN$ that can be reached from
	$v$ via a tree-path. Thus, $\cN$ is not tree-child; a contradiction.
	Since, by assumption, $\cN$ displays both $\cT$ and $\cT'$,
	Lemma~\ref{display} implies that
	$\gamma(v) \le 2$ for all $v \in V(\cN)-\{\rho_{\cN}\}$. Thus,
	$\gamma(v)=2$ must hold for all $v \in V(\cN)-\{\rho_{\cN}\}$.
	
	(ii) $\Rightarrow$ (iii) This is trivial.
	
	(iii) $\Rightarrow$ (i) By Lemma~\ref{display} it suffices to show that
	$\gamma_{\psi}(v)\leq 1$ and $\gamma_{\psi'}(v)\leq 1$ holds
	for all $v\in V(\cN)$. Assume for contradiction that there exists some
	$v\in V(\cN)$ and some tree in $ \{\cT,\cT'\}$, say $\cT$, such that 
	$\gamma_{\psi}(v)\geq 2$. Then $v\not=\rho_{\cN}$.
	In view of the assumptions on $\cN$ and $\cT'$, the last statement in 
		Lemma~\ref{display} implies that
	there must be a directed path in $\cT'$ that starts at
	$\rho_{\cT'}$ 
	such that the image under $\psi'$ of the last edge in this path
	passes through or ends at $v$. Hence, $\gamma_{\psi'}(v)\geq 1$
	must hold too. Thus, $\gamma(v)\geq 3$. By assumption, it follows that
	$v$ must be a tree vertex of $\cN$. In view of  Lemma~\ref{helper}(i),
	there must exist some vertex $w\in Ret(\cN)$ that is an ancestor of $v$.
	Without loss of generality, we may assume that $w$ is such that
	no vertex in $V(\cN)$ distinct from $w$ that is  above $v$ and
	below $w$ is contained in
	$Ret(\cN)$. By assumption,  it follows that $\gamma(w)=2$; a contradiction
	to the choice of $w$ and the fact that $\gamma(v)\geq 3$.
	\qed
	
	Motivated in part by this lemma, we
	say that a phylogenetic network $\cN$ on $X$
	\emph{rigidly displays} two phylogenetic trees $\cT$ and $\cT'$
	on $X$ if $\cN$ weakly displays
	$\cT$ and $\cT'$ via display maps $\psi$, $\psi'$ respectively, and, 
		for all $v \in Ret(\cN)$ we have $\gamma_{\psi,\psi'}(v) \le 3$
		and, for each parent $w\in V(\cN)$ of $v$, we have
		$\gamma_{\psi,\psi'}(w) \le 2$. For example, the
	network $\cN'$ pictured in Figure~\ref{fig:beaded} rigidly displays the two
	phylogenetic trees depicted in that figure. 
	
	Note that, in contrast to the definitions of displaying
		and weakly displaying which refer to a single tree, rigidly 
		displaying always refers to two trees. In addition, 
		by Lemma~\ref{display-gamma} it follows that if $\cT$
		and $\cT'$ are two phylogenetic trees on $X$ that are
		displayed by a temporal tree-child
		network $\cN$ on $X$, then $\cN$ also rigidly displays
		$\cT$ and $\cT'$.  We also have the following:
	
	\begin{lemma}\label{rigidbound}
		Suppose $\cN$ is a temporal tree-child network on $X$ 
		and that $\cN$
		rigidly displays two phylogenetic trees $\cT$ and $\cT'$ on $X$ via display maps
			$\psi$ and $\psi'$. Then
		$2 \le \gamma(v) \le 3$ for all $v \in V(\cN)-\{\rho_{\cN}\}$.	
	\end{lemma}
	\noindent {\em Proof:}
	Suppose $v \in  V(\cN)-\{\rho_{\cN}\}$.
	Then since $\cN$ rigidly displays $\cT$ and $\cT'$
	it also weakly displays $\cT$ and $\cT'$.
	Since $v\not=\rho_{\cN}$
	it follows that  $\gamma_{\psi}(v)\geq 1$ and that
	$ \gamma_{\psi'}(v)\geq 1$. 
	Hence,  $2 \le \gamma(v)$.
	
	For the remainder, assume for contradiction that there exists some
	$v\in  V(\cN)-\{\rho_{\cN}\}$ such that $\gamma(v) \ge 4$.
	Then $v$ must be a tree vertex of $\cN$ as $\cN$ rigidly displays
	$\cT$ and $\cT'$ and $\gamma(\rho_{\cN})=0$.
	Let $P: v_1,v_2,\dots,v_k,v$ be a longest directed path of tree vertices
	in $\cN$ that ends at $v$. Note 
	that $\gamma(v_i)\geq\gamma(v)$, for all $1\leq i\leq k$.
	Also note that since $\rho_{\cN}$ is not a tree vertex of $\cN$,
	we cannot have $v_1= \rho_{\cN}$.
	Let  $w\in V(\cN)$ denote the parent of $v_1$.
	Note that $\gamma(w) \ge 4$. Hence, we cannot have $w=\rho_{\cN}$.
	Since $\cN$ rigidly displays $\cT$ and $\cT'$
	it follows that $w$ must be a tree vertex of $\cN$. But then the extension
	of $P$ by $w$ results in a directed path of tree vertices of $\cN$
	that ends in $v$
	and that is longer than $P$; a contradiction.
	\qed	
	
	
	\begin{figure}[t]
		\center
		\scalebox{0.99}{\includegraphics{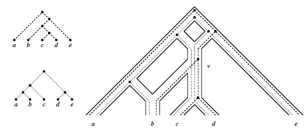}}
		\caption{The two phylogenetic trees on the left are weakly displayed
			by the network depicted on the right. However, they are not
			rigidly displayed by that network because
			$\gamma(v)=3$ and $v$ is the parent of a reticulation vertex.}
		\label{fig:example}
	\end{figure}
	
	Note that the converse of the last lemma does not hold in general  (see e.\,g.\,Figure \ref{fig:example}).
		We conclude this section with one more lemma that will be useful later.
	
	\begin{lemma} \label{split}
		Suppose that $\cN$ is a tree-child network on $X$ that
		rigidly displays two phylogenetic trees $\cT$
                and $\cT'$ on $X$, and
		that $\cT$ is weakly displayed via the display
                map $\psi$. 
		If $e=(u,v)$  is an edge of $\cT$
		such that $\psi[e]$
		passes through a vertex $w \in Ret(\cN)$,
		then $\psi(u)$ must be a parent of $w$ in $\cN$.
	\end{lemma}
	\noindent {\em Proof:}
	Suppose $e=(u,v)$ in $\cT$ is such that $\psi[e]$ passes
	through a vertex $w \in Ret(\cN)$. Assume for contradiction
	that $\psi(u)$ is not a parent of $w$.
	Let $p$ be the parent of $w$ in $\cN$ such that
	$p$ lies on $\psi[e]$. Then $\psi(u) \neq p$.
	As $\cN$ is tree-child, there must be a tree-path in $\cN$  starting at $p$ 
	and ending at some leaf $x \in X$. So, as $x$ is a leaf of $\cT$ and
	$\psi_{\cT}(u) \neq p$, 
	there must be some edge $e' \neq e$ in 
	$\cT$ such that $\psi[e']$ passes through $p$. Moreover,
	considering the leaf $x$ again, there
	must be an edge in $\cT'$ which maps to a path in $\cN$ via $\psi$ that
	either ends at or passes through $p$. 
	It follows that $\gamma(p) \ge 3$; a contradiction as
	$\cN$ rigidly displays $\cT$ and $\cT'$ and $p$ is the parent
	of a vertex in $Ret(\cN)$.
	\qed
	
	Note that as the example of the two phylogenetic trees and the
	network $\cN'$ in Figure~\ref{fig:nonexample} shows, the assumption that $\cN$
	is tree-child is necessary for Lemma~\ref{split} to hold. 
	
	\section{Fork operations}\label{sect:fork}
	
	In the next section we shall characterize when 
		two trees are rigidly displayed by a temporal tree-child network in terms
		of sequences of certain operations on these trees. The basis for these sequences
		are fork-operations which we shall now introduce.
	
	By a {\em fork} we mean a 2-leaved tree (i.\,e.\,a cherry),
	a 3-leaved rooted tree (a {\em 3-fork}) or a 4-leaved fully-balanced 
	rooted tree (a {\em 4-fork}). The following basic 
		fact concerning forks is straight-forward to show.
	
	\begin{lemma}\label{fork}
		Suppose $\cT$ is a phylogenetic
		tree with $n\geq 3$ leaves. If $n=3$ then $\cT$
		is a 3-fork and if $n\geq 4$ then $\cT$
		must contain a pendant subtree that is either a
		3-fork or a 4-fork.
	\end{lemma}
	
\begin{figure}[t]
		\center
		\scalebox{0.99}{\includegraphics{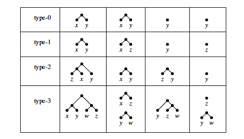}}
		\caption{Four operations (type-0, type-1, type-2, type-3)
			on two phylogenetic trees each applied to leaf $x$.
			In the 2nd and 3rd columns,
			subconfigurations of the two trees are represented (so,
			for example, for a type-0 operation, both trees have
			cherry $\{x,y\}$). In the 4th and 5th columns the result of applying
			the operation $o(x)$ is pictured (so, for example,
			applying a type-2 operation to the 3-fork and the cherry
			in row 3 results in a phylogenetic tree with cherry $\{z,y\}$
			and a phylogenetic tree  with the cherry $\{x,y\}$ replaced by $y$).
		}
		\label{fig:operations}
	\end{figure}

	A {\em fork-operation} $o=o(x)$
	for a pair of phylogenetic trees $\cT$ and $\cT'$ on $X$
	consists of a
	leaf $x \in X$, together with a fork in each of $\cT$ and $\cT'$ containing $x$
	as depicted in the second and third columns of Figure~\ref{fig:operations}.
	In case the type of fork-operation is relevant to the discussion
	we also write $o_i(x)$, $i\in\{0,1,2,3\}$,
	where $o_i$ is a type-$i$ operation.
	In addition,  we shall call $x$ the leaf {\em associated} to the operation.
	When we {\em apply} an operation $o$ to some element $x\in X$,
	we remove the leaf $x$ from both trees, and suppress
	any resulting vertices of degree 2 (removing the root and both
	edges incident with it in case $|X|=2$).
	
	Now, given two phylogenetic trees $\cT$ and $\cT'$ on the set
	$X=\{x_1,\ldots,x_m\}$, $m\geq 3$, we call  a sequence
	$(o(x_1),o(x_2),\dots,o(x_l))$ of $l$, $1 \le l \le m-2$
	fork-operations a {\em special sequence} for $\cT$ and $\cT'$
	if $o(x_l)$ is a type-1 operation on $\cT|_{X- \{x_1,\dots,x_{l-1}\}}$
	and $\cT'|_{\{X - \{x_1,\dots,x_{l-1}\}}$ and, in case $l>1$, the following
	properties hold:
	\begin{enumerate}
		\item[(i)] There exists some $\cT^* \in \{\cT,\cT'\}$ such that
		each $o(x_i)$, $1 \le i \le l-1$ is a type-2 or a type-3 operation
		applied to $x_i$
		and the associated 3- or 4-fork is a pendant subtree of
		$\cT^*|_{X-\{x_1,\dots,x_{i-1}\}}$, 
		\item[(ii)] the last-but-one operation $o(x_{l-1})$
		is a type-2 operation
		with fork $(p,(x_{l-1},x_l))$ and cherry $(x_{l-1},p)$ some
		$p\in X-\{x_1,\ldots, x_{l-1}\}$, and the
		last operation $o(x_l)$ is a type-1 operation
		with cherries $(p,x_l)$ and $(q,x_l)$, some
		$p,q\in X-\{x_1,\ldots, x_{l-1}\}$ distinct, and 
		\item[(iii)] if $l>2$ then
		$ lca_{\cT^*}(x_{l-1},x_l)\preceq lca_{\cT^*}(x_i,x_{l-1},x_l)$ must hold
		for all $1 \le i \le l-2$  for the tree $\cT^*$ in (i).
	\end{enumerate}
	
	To illustrate this definition, consider the phylogenetic
	network on $X=\{x_1,x_2,\ldots,x_6\}$  depicted in
	Figure~\ref{fig:example2}. Then $(o_3(x_5), o_2(x_3), o_1(x_4))$ is a
	special sequence for the two phylogenetic trees on $X$ also pictured in that
	figure where, for example, $o_3(x_5)$ is a fork-operation of type-3
	and the tree with cherry $\{x_4,x_5\}$
	is the tree $\cT^*$ mentioned in the definition.
	Note that an application of a special sequence always
	results in phylogenetic trees with at least two leaves.
	The following  proposition will be key to the proof of our
        main results.
	
	\begin{proposition}\label{special}
		Suppose that $\cN$ is a temporal tree-child network on $X$, $|X|\geq 3$,
		that rigidly displays two phylogenetic trees $\cT$ and $\cT'$ on $X$.
		If no type-0 operation can be applied to $\cT$ and $\cT'$,
		then there is a special sequence $\sigma$ for $\cT$ and $\cT'$.
		Moreover, the two phylogenetic
		trees resulting from applying $\sigma$
		can be rigidly displayed 
		by a temporal tree-child network $\cN'$ with
		$h(\cN')=h(\cN)-1 \geq 0$. 
	\end{proposition}
	\noindent {\em Proof:}
	Note first that $h(\cN)>0$ as otherwise $\cN$ would be
	a phylogenetic tree that is isomorphic with both
	$\cT$ and $\cT'$ implying that a type-0
	operation can be applied to $\cT$ and $\cT'$; a contradiction.
	
	Now, let $t:V(\cN) \to \mathbb R^{\geq 0}$ denote a temporal
        labelling for
	$\cN$ and pick some $v \in Ret(\cN)$
	whose value is maximum under $t$. 
	Let $u$ and $w$  be the parents of $v$.
	Note that $u\not=w$ as $\cN$ does not
	contain parallel edges. Also note that since $\cN$ does not
	contain shortcuts as it is normal, $u$ cannot be an ancestor of $w$ 
	and $w$ cannot be an ancestor of $u$.
	In particular, this implies that $u$ and $w$ must 
	be tree vertices. Let $p$ be the child of $u$ that is 
	not $v$ and, similarly, let $q$ be the child of $w$ that 
	is not $v$. We claim that $p$  is a leaf of $\cN$. 
	
	To see that this claim holds, assume
	for contradiction that $p$ is the root of a pendant subgraph
	$\cN^*$ of $\cN$. Note that the choice of $v$ implies that
	$\cN^*$ is in fact a pendant subtree of $\cN$. Moreover,
	Lemma~\ref{rigidbound} implies that there are display 
	maps for $\cT$ and $\cT'$ in $\cN$ such that $\gamma(u)=2$.
	By Lemma~\ref{helper}(ii) it follows that $\cN^*$ is a pendant
	subtree of both $\cT$ and $\cT'$. Hence,
	$\cT$ and $\cT’$ have a common cherry and, so, we can
	apply a type-0 operation to $\cT$ and $\cT'$; a contradiction. 
	Thus  $p$ must be a leaf of $\cN$. Applying similar arguments
	to $q$ implies that $q$ must also be a leaf of $\cN$.
	
	
	Since $\cN$ is temporal, the choice of $v$ implies
	that the child $s$ of $v$ is a leaf of $\cN$ or the
	root of a pendant subtree  of $\cN$.
	Assume first that $s$ is a leaf of  $\cN$.
	Then since $\cT$ and $\cT'$ are rigidly 
	displayed by $\cN$ and $\cT$ and $\cT'$ do not contain a common cherry,
	it is straight-forward to see using Lemma~\ref{split}
	that, without loss of generality, $\cT$ and $\cT'$ must contain the cherries 
	$\{p,s\}$ and $\{q,s\}$, respectively. Hence we can apply a
	type-1 operation to $s$.
	This gives a special sequence of length 1 for $\cT$ and $\cT'$,
	from which the first part of the proposition follows.

	So, suppose that $s$ is the root of a pendant subtree $\cT^*$
	of $\cN$, so that $\cT^*$ has at least 
	two leaves. Note first that $\gamma(v)=3$. Indeed since
	$\cT$ and $\cT'$ are rigidly displayed by $\cN$ we obtain
	$2\leq \gamma(v)\leq 3$ in view of Lemma~\ref{rigidbound}.
	If $\gamma(v)=2$ held then $\cT$ and $\cT'$ would
	have a common cherry which implies that
	a type-0 operation can be
	applied to $\cT$ and $\cT'$; a contradiction.
	We can therefore assume without loss of generality that $\cT^*$ is
	a pendant subtree of $\cT$, and that this tree together with the leaf $p$
	also forms a pendant subtree of $\cT$.
	
	In case $\cT^*$ has only two leaves $x$ and $y$, say, then since
	$\gamma(v)=3$ it follows that $\cT$ contains
	the 3-fork $(p,(x,y))$ and $\cT'$ contains, without loss of generality,
	the cherries $\{p,y\}$ and $\{q,x\}$.
	Hence we can apply a type-2 operation to $y$ and
	then apply a type-1 operation to $x$
	(since $\cT|_{X-\{y\}}$ and $\cT'|_{X-\{y\}}$ must
	contain the cherries $\{p,x\}$ and $\{q,x\}$, respectively).
	This gives a special sequence of length 2, from which the first part of the
	proposition again follows.

	Assume for the remainder that $\cT^*$ has at least three leaves. We claim
	that we can perform a sequence of type-2 and type-3 operations
	involving the removal 
	of an element from $L(\cT^*)$ one at a time, and at no stage 
	creating a common cherry, followed by a type-1 operation
	which,  when applied, results in a special sequence for $\cT$ and $\cT'$.
	We prove the claim by induction on the number $k \ge 2$ of leaves
	of $\cT^*$. Note that we have just shown that the claim holds
	for the base case $k=2$.
	So suppose the claim holds for all $k$, $k \ge 2$, and that $\cT^*$
	contains $k+1$ leaves. Note that as $k+1 \ge 3$, Lemma~\ref{fork}
	implies that $\cT^*$ contains either a 3-fork or a 4-fork. 
	
	Suppose $\cT^*$ contains a 3-fork $t=(a,(b,c))$ where $a,b,c\in X$
	are distinct.
	Then $t$ must be a pendant subtree of $\cT$. As $\cT$ and $\cT'$
	have no cherries
	in common and $\gamma(v)=3$, it follows that we may assume
	without loss of generality that $\cT'$
	contains the cherry $\{a,c\}$. 
	Hence, we can apply the type-2 operation $o(c)$.
	Note that this creates a cherry $\{a,b\}$ in $\cT|_{X-\{c\}}$
	which is not a cherry in $\cT'|_{X-\{c\}}$.
	Moreover, by induction we obtain a special sequence
	$\sigma_c=(o(x_1), o(x_2),\ldots, o(x_l))$, $1\leq l\leq n-2$
	for $\cT|_{X-\{c\}}$ and $\cT'|_{X-\{c\}}$. Put $x=x_l$
	and, if $l\geq 2$, put $y=x_{l-1}$. Note that if $l=1$, then
	$(o(c), o(x_1))$ is a special sequence for $\cT$ and
	$\cT'$. So assume $l\geq 2$. If 
	$lca_{\cT}(x,y)\preceq lca_{\cT}(x,y,c)$ then $(o(c),\sigma_c)$
	is clearly a special sequence for $\cT$ and $\cT'$. And if
	$lca_{\cT}(x,y,c)\prec lca_{\cT}(x,y)$ then $\sigma_c$ is also a special 
	sequence for $\cT$ and $\cT'$ since  applying the operation
	$o(c)$ to $\cT$ and $\cT'$ does not affect any of the operations 
	$o(x_i)$, $1\leq i\leq l$.

	Suppose $\cT^*$ contains a 4-fork $t=((a,b),(c,d))$ where $a,b,c,d\in X$
	are distinct.
	Then $t$ must again be a pendant subtree of $\cT$. As $\cT$ and $\cT'$
	have no cherries in common and $\gamma(v)=3$,
	it follows that, as before, we may assume
	without loss of generality that $\cT'$ contains the cherries $\{a,c\}$
	and $\{b,d\}$. 
	Hence, we can perform the type-3 operation $o(c)$.
	Note that this creates a 3-fork $(d,(a,b))$ in $\cT|_{X-\{c\}}$
	and that $\{a,b\}$ is not a cherry in $\cT'|_{X-\{c\}}$.
	Moreover, by induction, we obtain a special sequence $\sigma_c$ for
	$\cT|_{X-\{c\}}$ and $ \cT'|_{X-\{c\}}$. Let $x$ denote the leaf
	to which the (sole) type-1 operation is applied
	and let $y$ denote the leaf to which the last type-2 operation is applied. 
	Then similar arguments as in the previous case imply
	that $(o(c),\sigma_c)$
	is  a special sequence for $\cT$ and $\cT'$ in case
	$lca_{\cT}(x,y)\preceq lca_{\cT}(x,y,c)$ and that 
	$\sigma_c$ is a special
	sequence for $\cT$ and $\cT'$ otherwise.
	This concludes the proof of the induction step and, therefore,
	the proof of the claim. This completes again the proof of the first
	part of the proposition.
	
	To complete the proof, note that as $\cT^*$ is a pendant subtree of
	$\cN$, we can remove  $\cT^*$ and $v$ (plus all its
	incident edges) from $\cN$, and suppress
	the resulting vertices of degree two to obtain a network $\cN'$ with 
	$h(\cN')=h(\cN)-1 \geq 0$. As $\cN$  
	rigidly displays $\cT$ and $\cT'$, it follows
	that $\cN'$ rigidly displays their restrictions
	$\cT|_{X-L(\cT^*)}$ and $\cT'|_{X-L(\cT^*)}$.
	Moreover, as $\cN$ is tree-child and $v\in Ret(\cN)$,
	we have that $\cN'$  is also tree-child. Since
	$p$ and $q$ are leaves of $\cN$ and $\cN$ is temporal,
	it follows that $\cN'$ is temporal.
	\qed

	\section{Fork-picking sequences}\label{sect:forkpicking}
	
	In this section we characterize when 
		two trees are rigidly displayed by a temporal tree-child network, in terms of
		a generalization of special sequences which we now introduce.
	Suppose that $\cT$ and $\cT'$ are two phylogenetic trees
		on $X=\{x_1,\ldots, x_n\}$ where $n\geq 2$ and that
		$\sigma=(o(x_1),o(x_2),\ldots,o(x_{n-1}))$
		is a sequence of fork-operations for $\cT$ and $\cT'$.
		Then we call $\sigma$ a {\em fork-picking sequence for $\cT$ and $\cT'$} if 
		$\sigma$ is of the form $(C_1,S_1,C_2,S_2,\dots, C_k,S_k,C_{k+1})$,
		some $k\geq 0$, such that
		\begin{itemize}
			\item[(i)] for all $1\leq i\leq k+1$, we have that
			$C_i$ is a (possibly empty, except in case $i=k+1$)
			sequence of solely type-0 operations for $\cT$ and $\cT'$, and
			\item[(ii)] for all $1\leq i\leq k$, $S_i$ is a special sequence 
			for $\cT|_Y$ and $\cT'|_Y$, where 
			$Y=\{y_1 = x_p, y_2 = x_{p+1}, \dots, y_{n-p+1}= x_n\}$ and
			$o(x_p)$ is the first operation in $S_i$ (so that, in particular, $n-p+1\ge 3$). 
		\end{itemize}
		To ease readability, we omit all those $C_i$ that are empty
		when writing down fork-picking
		sequences. Note that it follows from the definition that any fork-picking sequence
		can be decomposed in a unique way into the 
		form $(C_1,S_1,C_2,S_2,\dots, C_k,S_k,C_{k+1})$, 
		and that all of the subsequences $S_i$ 
		are non-empty.
	
\begin{figure}[t]
		\center
		\scalebox{0.99}{\includegraphics{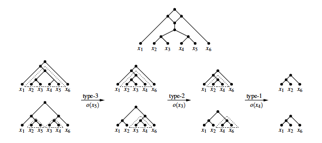}}
		\caption{For the two pictured phylogenetic trees $\cT$ and $\cT'$ on
			$X=\{x_1,\ldots, x_6 \}$ and the phylogenetic network $\cN$ on
			$X$ that rigidly displays them, we depict
			a fork-picking
			sequence $(S_1,C_2)$ for $\cT$ and $\cT'$ which has
			weight 1. In that sequence  $S_1$ is
			the indicated special sequence, and $o_0(x_1)$
			and $o_0(x_2)$ make up $C_2$. The forks in $\cT$ and $\cT'$
			to which a fork-operation is applied is indicated by dotted
			triangles.
		}
		\label{fig:example2}
	\end{figure}

	To illustrate this definition, consider again
	the phylogenetic network on $X=\{x_1,\ldots, x_6\}$
	pictured in Figure~\ref{fig:example2}. Then 
	$\sigma^*=(o_3(x_5),o_2(x_3), o_1(x_4),o_0(x_1), o_0(x_2))$
	is a fork-picking
	sequence for the two phylogenetic trees also depicted in
	that figure, since it is of the form $(S_1,C_2)$
		where $S_1 = (o_3(x_5),o_2(x_3), o_1(x_4))$ is the special sequence for $\cT$ and
		$\cT'$ considered in the previous section, $C_2= (o_0(x_1), o_0(x_2))$,
		and $C_1$ is the empty sequence. 
	
	We now provide a link between $h(\cN)$ for a temporal
		tree-child network $\cN$ that rigidly displays two trees and 
		fork-picking sequences for these trees.
		We define the {\em weight} $w(\sigma)$ of a fork-picking
		sequence $\sigma$ to be the number of 
		special sequences in $\sigma$ (or, equivalently, the number of
		type-1 operations in $\sigma$).
	
	\begin{theorem}\label{weighting}
		Suppose that $\cN$ is a temporal tree-child network on $X$
		that rigidly displays two phylogenetic trees $\cT$ and $\cT'$
		on $X$. Then there is a fork-picking sequence $\sigma$
		for $\cT$ and $ \cT'$ with $h(\cN) \ge w(\sigma)$.
	\end{theorem}
	\noindent {\em Proof:}
	We prove the theorem by
		induction on $h(\cN)$.  If $h(\cN)=0$,
		then $\cN$, $\cT$, and $\cT'$ are all isomorphic to
		one another. But then we can take a fork-picking sequence
		$\sigma$ for $\cT$ and $\cT'$ consisting solely of type-0 operations
		(i.e. $\sigma=(C_1)$), and so $w(\cN)=0=h(\cN)$.
	
	Now, assume that $h(\cN)=k$, some $k>0$, and that
		the theorem holds for all
		temporal tree-child networks $\cN'$ with $0\leq h(\cN')<k$.
	
	Apply type-0 operations to $\cT$ and $\cT'$ until no
		more can be applied. If 
		this sequence $C_1$ of operations has length $|X|-1$, then it is
		a fork-picking sequence for $\cT$ and $\cT'$ and $w(\sigma)=0< k =h(\cN)$,
		and so the theorem holds.  Otherwise, 
		let $\cT_1$ and $\cT_1'$ be the phylogenetic trees resulting after applying the operations 
		in $C_1$, noting that $|\cL(\cT_1)|=|\cL(\cT'_1)| \ge 3)$.
	
	Since by construction no type-0 operation
		can be applied to $\cT_1$ and $\cT_1'$,
		by Proposition~\ref{special} it follows 
		that there is a special sequence $S_1$ 
		for $\cT_1$ and $\cT_1'$,
		and that the two phylogenetic
		trees $\cT_2$ and $\cT_2'$ resulting from applying $S_1$
		can be rigidly displayed 
		by a temporal tree-child network $\cN'$ with
		$h(\cN')=h(\cN)-1 \geq 0$.
	
	It follows by induction that there is a fork-picking
		sequence $\sigma'= (C_1',S_1',\dots,C'_{k'+1})$,
		some $k'\geq 1$, for $\cT_2$ and $\cT_2'$
		such that $k-1=h(\cN')\geq w(\sigma')=k'$.
		Hence, $\sigma= (C_1,S_1,C'_1,S_1',\dots,C'_{k+1})$ is a
		fork-picking sequence for $\cT$ and $ \cT'$ such
		that $h(\cN)=k\geq w(\sigma)$.
	\qed
	
	Now, as defined in \cite{HLS13}, we say that an ordering
		$\sigma = (x_1 , x_2, \ldots, x_{n-1},x_n)$ of $X$
		is a {\em cherry-picking sequence} for
		two phylogenetic trees $\cT$ and $\cT'$ on $X$ if
		for all $1\leq i\leq n-1$, $x_i$ is contained in a cherry
		in both $\cT_i=\cT|_{X-\{x_1,\ldots, x_{i-1}\}}$ and
		$\cT_i'=\cT'|_{X-\{x_1,\ldots, x_{i-1}\}}$. 
		In addition, the {\em cherry-count $c_i(\sigma)-1\in \{0,1\}$ associated to $x_i$}
		is 1 if the cherries in $\cT_i$ and $\cT'_i$ containing $x_i$
		are different and 0 else.
	
	Note that every cherry-picking sequence
		$\sigma=(x_1,x_2,\dots,x_n)$ for two phylogenetic
		trees $\cT$ and $\cT'$ gives rise to a
		fork-picking sequence for $\cT$ and $\cT'$. Namely, we
		make a sequence of operations $(o(x_1),\dots,o(x_{n-1}))$ 
		with a type-1 operation applied to $x_i$ if $c_i(\sigma)=1$ and a
		type-0 operation applied to $x_i$ if $c_i(\sigma)=0$.
		In addition, any fork-picking sequence
		$(o(x_1),\dots,o(x_{n-1}))$ for two phylogenetic trees
		$\cT$ and $\cT'$ on $X$ clearly gives rise to 
		the cherry-picking sequence $(x_1,x_2,\dots,x_n)$. 
		For example, the cherry-picking sequence
		$(x_5,x_3,x_4,x_2,x_1)$ with cherry counts
		$(1,1,1,0,0)$ arises from the fork-picking sequence $\sigma^*$ 
		given above for the two trees in Figure~\ref{fig:example2}.
		Using these observations we obtain the following result.
	
	\begin{corollary}\label{rigid-cherry}
		Suppose that $\cT$ and $\cT'$ are two phylogenetic trees on $X$.
		Then the  following statements are equivalent:
		\begin{itemize}
			\item[(i)] $\cT$ and $\cT'$ are rigidly displayed by a
			temporal tree-child network on $X$.
			\item[(ii)] $\cT$ and $\cT'$ are displayed by a
			temporal tree-child network on $X$.
			\item[(iii)] there is a cherry-picking sequence for
			$\cT$ and $\cT'$
			\item[(iv)] there is a fork-picking sequence for
			$\cT$ and $\cT'$
		\end{itemize}
	\end{corollary}
	\noindent {\em Proof:}
	(ii) $\Rightarrow$ (i) If two phylogenetic
	trees are displayed by a phylogenetic network then they are rigidly 
	displayed by that network.
	
	(iii) $\Rightarrow$ (ii)  Apply \cite[Theorem 1]{HLS13}, which states
		that two phylogenetic trees are displayed by a  temporal tree-child network
		if and only if there is a cherry-picking sequence for them.
	
	(i) $\Rightarrow$ (iv) Apply Theorem~\ref{weighting}.
	
	(iv) $\Rightarrow$ (iii) Apply the observation stated before the statement 
	of the corollary i.\,e.\,that a fork-picking sequence gives rise to 
	a cherry-picking sequence.
	\qed
	
	Note that  the temporal tree-child networks whose existence is guaranteed
		in Corollary~\ref{rigid-cherry}(i) and (ii) need not be the same.
	
	Corollary~\ref{rigid-cherry} also sheds light on the following decision problem:
	
	\noindent {\sc Rigidly Displaying}\\
	\noindent{\bf Input}: Two
	phylogenetic trees $\cT$ and $\cT'$ on $X$.\\
	\noindent{\bf Output}: Does there exist a
	temporal tree-child network on $X$
	that rigidly displays $\cT$ and $\cT'$?
	
	\noindent Indeed, Corollary~\ref{rigid-cherry} and 
		the main result in \cite[Thoroem 1]{Dl17} (which states that 
		it is NP-complete to decide whether or not there is a cherry-picking 
		sequence for two phylogenetic trees) immediately imply:
	
	\begin{corollary}
		The decision problem {\sc Rigidly Displaying} is NP-complete.
	\end{corollary}
	
	
	\section{A characterization of the rigid hybrid number of two trees}\label{sect:characterize}

	In this section we show that in
		case two phylogenetic trees can be 
		rigidly displayed by a temporal tree-child network, then 
		their rigid hybrid number is equal to the weight of a minimum weight fork-picking 
		sequence for the two trees.
	
	To this end, if
	there is some fork-picking sequence for two phylogenetic trees
	$\cT$ and $\cT'$ on $X$ (or equivalently by Corollary~\ref{rigid-cherry}, $\cT $ and $ \cT'$ are
		rigidly displayed by some temporal tree-child network on $X$), 
	we  define 
	$$
	s_r(\cT,\cT') = \min\{w(\sigma) \,:\, \sigma \mbox{ is a
		fork-picking sequence for } \cT \mbox{ and } \cT'\},
	$$
	and 
	$$
	h_r(\cT,\cT') = \min\{h(\cN) \,:\, \cN \mbox{ is temporal tree-child
		and rigidly displays } \cT \mbox{ and }\cT'\}.
	$$
	
\begin{figure}[t]
		\center
		\scalebox{0.99}{\includegraphics{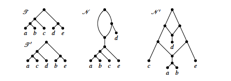}}
		\caption{Two phylogenetic trees $\cT$ and $ \cT'$ on
			$X=\{a,\ldots, e\}$ whose 
			weak hybrid number $h_{wd}(\cT,\cT')$ is 1 (see $\cN$ for
			a phylogenetic network that weakly displays both trees), whose
			rigid hybrid number is
			$h_r(\cT,\cT')=2$ (see $\cN'$ for a temporal tree-child
			network which rigidly displays $\cT$ and $\cT'$; the fact 
			that $h_r(\cT,\cT')\neq 1$ can be proven using
			similar arguments to those
			used in the proof of Proposition~\ref{nastyexample}),
			and whose hybrid number $h(\cT,\cT')=2$. }
		\label{fig:weakdiffer}
	\end{figure}

	We call $h_r(\cT,\cT')$ the {\em rigid (temporal tree-child)
			hybrid number} for  $\cT$ and $\cT'$. Note
		that in case this number exists it can be different 
		from the weak hybrid number, which must also 
		exist (e.g. see Figure~\ref{fig:weakdiffer}).
	
	\begin{theorem}\label{bounded}
		Suppose  $\cT$ and $\cT'$ are two phylogenetic trees on $X$
		and that
		$\sigma$ is a fork-picking sequence  for $\cT$ and $ \cT'$.
		Then there exists a temporal tree-child network $\cN_{\sigma}$
		on $X$ which 
		rigidly displays $\cT$ and $\cT'$ and such that
		$w(\sigma) \ge h(\cN_{\sigma})$.
	\end{theorem}
	\noindent {\em Proof:}
	We establish the theorem using induction on $w(\sigma)$.
	
	If $w(\sigma)=0$ then $\sigma =(C_1)$,
	where $C_1$ consists solely of type-0 operations. 
	Hence $\cT$ and $\cT'$ are isomorphic and the
	required temporal tree-child network $\cN_{\sigma}$ is 
	given by $\cT$.
	
	Now suppose that $\sigma$ is a fork-picking sequence 
		for $\cT$ and $\cT'$ with weight $w(\sigma)=k$, some $k>0$, 
		and that the theorem holds for all fork-picking sequences $\sigma'$ 
		with $w(\sigma')< k$.
	
	As $w(\sigma)=k$, $\sigma$ is of the form
		$(C_1,S_1,C_2,S_2,\dots,S_k,C_{k+1})$. 
		Let $Y$ be the set of elements $y$ in $X$ such 
		that $o(y)$ is {\em not} in the sequence $(C_1,S_1)$. 
		Then, as $C_{k+1}$ is not the empty sequence, 
		$\sigma'= (C_2,S_2,\dots,S_k,C_{k+1})$
		is a fork-picking sequence for  $\cT|_Y$ and $\cT'|_Y$, 
		with $w(\sigma')=k-1$.
		By induction, it follows that there is a temporal
		tree-child network $\cN'$
		with $h(\cN') \le w(\sigma')$
		that rigidly displays $\cT|_Y$ and $\cT'|_Y$. Let $t'$ denote a
		temporal labelling for $\cN'$.
	
	We now construct a temporal tree-child network $\cN_{\sigma}$ from $\cN'$.
		We first consider the case that $(C_1)$ is the empty sequence.
		Let $Y_1=\{y_1,\ldots, y_l\}$, $l\geq 1$, be such 
		that  $S_1=(o(y_1),\dots,o(y_l))$.  To ease notation, put $x=y_l$ 
		and, if $l\geq 2$, $y= y_{l-1}$.
	
	Since $o(x)$ is a type-1 operation, there exist
		$p\not=q\in Y$ such that, without loss of generality, $\{p, x\}$
		is a cherry in $\cT|_{Y\cup\{x\}}$ and
		$\{q, x\}$ is a cherry in $\cT'|_{Y\cup\{x\}}$.
	Subdivide the pendant edges in $\cN'$ incident with $p$ and $q$ 
	by adding two new vertices $u_p$ and $u_q$, respectively. 
	Also, add in the leaf $x$ below a newly added reticulation vertex $v$ 
	which has parents $u_p$ and $u_q$. Denote the resulting
	phylogenetic network by $\cN''$. Note that since $\cN'$
	is tree-child we also have that $\cN''$ is tree-child.
	Set $t(v)=t(u_p)=t(u_q)$
	where $t(u_p)=t(u_q)$ is chosen appropriately so that, together with $t'$,
	we obtain a temporal labelling $t''$ for $\cN''$.  
	Clearly, $\cN''$ rigidly displays
	$\cT|_{Y\cup\{x\}}$ and  $\cT'|_{Y\cup\{x\}}$.
	Putting $\cN_{\sigma}=\cN''$ this completes the
	proof of the theorem in case $l=1$ since
	$h(\cN'')=h(\cN')+1\leq w(\sigma')+1 =w(\sigma)$.
	
	Assuming $l \ge 2$, we now insert $y$ into $\cN''$. By 
	definition of a special sequence, $o(y)$ is a type-2 operation.
	Without loss of generality, we may assume that $\cT|_{Y\cup\{x,y\}}$
	has a fork $(p,(x,y))$ and that $\cT'|_{Y\cup\{x,y\}}$ has a cherry $\{y,p\}$
	some $p\in X-\{x,y\}$. Then we can construct a new temporal tree-child 
	network $\cN'''$ which rigidly displays $\cT|_{Y\cup\{x,y\}}$ and
	$\cT'|_{Y\cup\{x,y\}}$ by inserting a pendant edge containing
	$y$ into the pendant edge of $\cN''$ incident with $x$
	to form a cherry $\{x,y\}$ in $\cN''$
	and defining the temporal labelling $t'''$ of $\cN$ appropriately using $t''$. 
	Note that $h(\cN''')=h(\cN'')$ and so the theorem is also proven for
	$\cN_{\sigma}=\cN''$ if $l=2$.
	
	Assume $l\ge 3$. Bearing in mind Property~(iii) of
	a special sequence, suppose we have created a temporal tree-child
	network $\cN^*$ from $\cN'''$
	by successively inserting, for all $i+1 \le j \leq l-2$,
		the leaves $y_j$ below the parent of the cherry $\{x,y\}$ in $\cN'''$
	to create a pendant subtree $\overline{\cT}$ 
	with leaf set $Y_2=\{y_{i+1},\dots,y_{l-2}, y,x\} \subseteq Y_1$ so that 
	$\cN^*$ rigidly displays the phylogenetic trees
	$\cT|_{Y\cup Y_2}$ and 
	$\cT'|_{Y\cup Y_2}$.
	Without loss of generality, we may assume that $\cT$ is the tree
	$\cT^*$ in the definition of a special sequence for $\cT$ and $\cT'$.
	
	Consider operation $o(y_i)$. Then, by definition of a special 
	sequence, $o(y_i)$ is either a type-2 operation or a type-3 operation, for 
	which the 3-fork and 4-fork, respectively, is a pendant subtree 
	of $\cT_1=\cT|_{Y\cup Y_2\cup\{y_i\}}$
	and $lca_{\cT}(y_i,y,x) \succeq lca_{\cT}(y,x)$.
	Put $\cT_1'=\cT'|_{Y\cup Y_2\cup\{y_i\}}$.
	
	If $o(y_i)$ is a type-2 operation, then let $(a,(y_i,b))$ denote the
	3-fork of $\cT_1$ where $a,b\in X-\{y_i\}$ distinct. Then
	$\{a,b\}$  must be a cherry in $\cT_1'$.
	Since $lca_{\cT}(y_i,x,y) \succeq lca_{\cT}(x,y)$ it follows
	by the choice of $\cT^*$ that $\{a,b\}$
	is a cherry in $\overline{\cT}$.
	We can therefore first
	insert $y_i$ into the pendant edge of $\cN^*$ incident with $b$ and
	then extend the temporal labelling of $\cN^*$ 
	so as to obtain a temporal tree-child network that 
	rigidly displays  $\cT_1$
	and $\cT'_1$.
	
	If $o(y_i)$ is a type-3 operation, then let
	$((a,y_i),(b,c))$ denote the 4-fork in $\cT_1$
	where $a,b,c\in X-\{y_i\}$ are pairwise distinct.
	Then $(a,(b,c))$ must be a 3-fork in
	$\cT_1|_{Y\cup Y_2}$.
	As $lca_{\cT}(y_i,y,x) \succeq lca_{\cT}(y,x)$
	the choice of $\cT^*$ implies that
	this 3-fork must be a pendant subtree of $\overline{\cT}$. 
	We can therefore first insert $y_i$ into the pendant edge of $\cN^*$
	incident with $a$ and then extend the temporal labelling of $\cN^*$ so as to
	obtain a temporal network that 
	rigidly displays  $\cT_1$ and $ \cT'_1$.
	
	In summary, we can insert all of the elements
	of $Y$ into $\cN'$ in this way until we obtain a temporal tree-child
	network $\cN_{\sigma}$ with $h(\cN_{\sigma})= h(\cN')+1$ which rigidly
	displays $\cT$ and $\cT'$. It follows by induction that 
	$$
	w(\sigma) = w(\sigma')+1 \ge h(\cN') +1 = h(\cN_{\sigma}),
	$$
	which completes the proof of the theorem in case $C_1$ is empty.
	
	If $C_1$ is not empty then we first insert all elements of $Y_1$ into
	$\cN'$ as described in the previous case to obtain a network $\cN_1$
	which rigidly displays $\cT|_{Y\cup Y_1}$ and $\cT'|_{Y\cup Y_1}$ and for which
	$w(\sigma) \geq h(\cN_1)$ holds. Into $\cN_1$ we then insert 
	all elements $z\in X$ for which $o(z)$ is contained
	in $(C_1)$ in any order to obtain a new temporal tree-child
	network $\cN_{\sigma}$  which rigidly
	displays $\cT$ and $\cT'$. Clearly, $h(\cN_{\sigma})= h(\cN_1)$.
	Since $\sigma''=(C_1,S_1,C_2,S_2,\dots,S_k,C_{k+1})$
	is a fork-picking sequence for $\cT$ and $\cT'$ and
	$w(\sigma)=w(\sigma'')$ the theorem holds in
	this case too.
	\qed
	
	As an immediate consequence of Theorems~\ref{weighting}
        and \ref{bounded},
		we have the following result.
	
	\begin{corollary}\label{rigidsize}
		If two phylogenetic trees $\cT$ and $ \cT'$
		on $X$ are rigidly displayed by some
		temporal tree-child network on $X$, then 
		$s_r(\cT,\cT')=h_r(\cT,\cT')$.
	\end{corollary}

	\section{The relationship between the temporal and rigid hybrid numbers}\label{sect:relationship}
	
	For two phylogenetic trees $\cT$ and $\cT'$ on $X$ 
		that can be displayed by some temporal tree-child network, 
		the {\em temporal hybrid number} of  $\cT$ and $\cT'$ \cite{HLS13} is defined as
		\begin{eqnarray*}
			h_t(\cT,\cT')
			=\min\{h(\cN) \,:\, \cN \mbox{ is a temporal tree-child
				network that displays } \cT \mbox{ and } \cT'\}.
		\end{eqnarray*}
		Note that in case this number exists, the 
		temporal hybrid number for the two trees is not necessarily equal 
		to their hybrid number \cite[Figure 1; also p. 1889]{HLS13}. 
	
	Now, given two phylogenetic trees $\cT$ and $\cT'$ on $X$,
		Corollary~\ref{rigid-cherry} implies that
		the temporal hybrid number $h_t(\cT,\cT')$ of $\cT$ and $\cT'$
		exists if and only if the rigid hybrid number $h_r(\cT,\cT')$ exists.
		Thus it is of interest to understand how the quantities $h_t(\cT,\cT')$
		and $h_r(\cT,\cT')$  are related to one another.
		Clearly, if these numbers both exist, then $h_r(\cT,\cT') \le h_t(\cT,\cT')$. 
		In this section, we show that the difference
		$h_t(\cT,\cT')-h_r(\cT,\cT')$  can grow as a linear function of $|X|$.
	
\begin{figure}[t]
		\center
		\scalebox{0.99}{\includegraphics{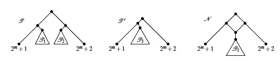}}
		\caption{Two phylogenetic trees $\cT$ and
			$\cT'$ on the set $X=\{1,2,\dots,2^m+2\}$ as defined in the text.
			Both are rigidly displayed by the depicted
			temporal tree-child network $\cN$ on $X$.}
		\label{fig:special}
	\end{figure}

\begin{figure}[t]
		\center
		\scalebox{0.99}{\includegraphics{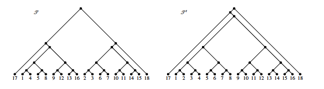}}
		\caption{The two phylogenetic
			trees $\cT$ and $\cT'$ in Figure~\ref{fig:special} for the
			case $m=4$.}
		\label{fig:differ}
	\end{figure}

	To this end, assume that $m \ge 3$.
	Consider the two phylogenetic trees
	$\cT$ and $\cT'$ on $X=\{1,\dots,2^m,2^m+1,2^m+2\}$,
	given in Figure~\ref{fig:special}. In that figure, $\cT_1$ and $\cT_2$ are 
	both fully balanced phylogenetic
	trees with $2^{m-1}$ leaves each (as indicted 
	in Figure~\ref{fig:differ} for the case $m=4$)
	and $\cT_3$ is a fully balanced phylogenetic
	tree with $2^m$ leaves (again as indicated in Figure~\ref{fig:differ} for the
	case $m=4$).
	In $\cT_3$, we label the pendant subtrees of size
	4 with the labels $\{4i-3,\dots,4i\}$, $1 \le i \le 2^{m-2}$
	as indicated in Figure~\ref{fig:differ},
	and in  $\cT_1$ and $\cT_2$ we label the pendant subtrees
	of size 4 by interchanging the labels
	as also indicated in Figure~\ref{fig:differ}.

	\begin{theorem}\label{big}
		For the two phylogenetic trees $\cT$ and $\cT'$ on $X=\{1,\dots,2^m+2\}$,
		$m\geq 3$,
		pictured  in Figure~\ref{fig:special}, we have
		$h_t(\cT,\cT')-h_r(\cT,\cT') \ge \frac{|X|}{4} - 3$.
	\end{theorem}
	\noindent {\em Proof:}
	First note that $h_r(\cT,\cT')=1$, as $\cT$ and $\cT'$
	are rigidly displayed by the temporal tree-child network $\cN$ pictured in 
	Figure~\ref{fig:special}.
	
	We now show that $h_t(\cT,\cT')\ge 2^{m-2}-1$ from
	which the theorem follows. First note that, by
	Corollary~\ref{rigid-cherry}, there must exist
	a temporal tree-child network that displays both
	$\cT$ and $\cT'$. By \cite[Theorem 3.3]{HL13},
		it follows that $h_t(\cT,\cT')$ is equal to the number of 
		components in a maximum temporal agreement
		forest for $\cT$ and $\cT'$ minus 1, where such a forest 
		is defined as follows. All phylogenetic trees considered
		in the definition are ``planted"  by adding a new root plus an 
		edge to their roots, and trees with one leaf are also allowed.
		A collection $\{\cT_0,\dots,\cT_k\}$ of
		$k=h_t(\cT,\cT')$ planted trees $\cT_i$ is 
		a maximum temporal agreement forest for 
		(planted versions of) $\cT$ and $\cT'$ if 
		the following three properties hold, where
		$ X_i:=L(\cT_i)$, $0\leq i\leq k$:
		\begin{itemize}
			\item [(P1)] The set $Z=\{X_0, \dots X_k\}$ is a
			partition of $X$.
			\item [(P2)] For all $0 \le i \le k$, $\cT_i \cong \cT|_{X_i}
			\cong \cT'|_{X_i}$.
			\item[(P3)] Denoting for $0\leq i\leq k$ the root of $\cT_i$ by $\rho_i$, 
			there exist injective maps $\chi: \{\rho_0,\rho_1,\ldots,\rho_k\} \rightarrow V(\cT)$ 
			and $\chi': \{\rho_0,\rho_1,\ldots,\rho_k\} \rightarrow V(\cT')$ 
			such that any two trees in $\{\cT(X_i\cup\{\chi(\rho_i)\}):0\leq i\leq k\}$ 
			and $\{\cT'(X_i\cup\{\chi'(\rho_i)\}):0\leq i\leq k\}$ are edge-disjoint rooted 
			subtrees of $\cT$ and $\cT'$, respectively. 
		\end{itemize}

	We now claim that for every set $\{4i-3,\dots,4i\}$,
	$1 \le i \le 2^{m-2}$, at least one of the sets 
	$\{4i-3\}$, $\{4i-2\}$, or $\{4i-3, 4i-2\}$
	must be contained in $Z$. This implies
	that $k+1 \ge 2^{m-2}$ from which the theorem immediately follows. 
	
	For simplicity, we prove the claim for the
	case $i=1$; the argument for the remaining cases $2 \le i \le 2^{m-2}$ is similar.
	
	We show that at least one of the subsets $\{1\}$, $\{2\}$ and $\{1,2\}$ of
	$\{1,\ldots,4\}$ is contained in
	$Z$. Suppose $\{1,2\}$ is not contained in $Z$.  
	Then, using Properties~(P1) -- (P3) it is straight-forward to check that 
	either  (a) $\{1\}$ and $\{2\}$ are both contained in $Z$, 
	(b) $\{1\} \cup Y$ and $\{2\}$ 
	are both contained in $Z$ for some non-empty $Y \subseteq X - \{1,2\}$ or
	(c) $\{1\}$, $\{2\} \cup Y$  are both contained in $Z$
	for some non-empty $Y \subseteq X - \{1,2\}$. Moreover, by
	Property~(P3), we cannot have that  $\{1,4\}$ and $\{2,3\}$,
	or $\{1,3\}$ and $\{2,4\}$ are contained in $Z$. Finally, since
	the restrictions of $\cT$ and $\cT'$ to $\{1,2,3\}$, or to
	$\{1,2,4\}$, or to $\{1,2\}\cup Y$, some non-empty set
	$Y\subseteq X-\{1,2,3,4\}$, are non-isomorphic, the
	claim follows.
	\qed
	
	\noindent {\bf Remark:} Note that the trees in Figure~\ref{fig:special} also provide an 
		example where $h_t(\cT,\cT') - h_{wd}(\cT,\cT')$  grows as a
		function of $|X|$ since $h_{wd}(\cT,\cT')=h_r(\cT,\cT')=1$.
	
	\section{Discussion}\label{sect:discussion}
	
	In this paper we have introduced the concept of a network 
	rigidly displaying a set of phylogenetic trees. We have shown
	that the rigid hybrid number is different from the beaded hybrid 
	number, and that it can be quite different from the 
	temporal hybrid number. We have also 
	characterized when two trees can be rigidly displayed 
	by a temporal tree-child network.
	
	There remain several open problems. First,
		it is well-known that the hybrid number is
		closely related to the size of a maximum agreement forest for 
		two phylogenetic trees \cite{B05}. It
		would therefore be of interest to know if there is some
		analogue of a maximum agreement forest for rigidly displaying two trees. Results
		in \cite{HL13}, including the one mentioned above, 
		concerning temporal agreement forests for two phylogenetic trees displayed
		by temporal tree-child networks could be useful for studying this question.
		In addition, it could be interesting to 
		define and study rigid hybrid numbers for three or
		more trees. For example, 
		we could try to understand $r$-rigidly displaying, where 
		$r$ is the maximum number of edges that come together at each 
		reticulation (note that in this paper
		we have investigated the concept of $r$-rigidly displaying for $r=3$).
		Recently, there has been 
		work on understanding the hybrid number for arbitrary sets of
		trees \cite{L19}  which might be relevant.
	
	
	More generally, several questions remain concerning
	the notion of weakly displaying. 
	For example,  it would be interesting to
	know how large the difference can potentially be
	between the hybrid number and the weak hybrid number
	for a collection of phylogenetic trees.  As this appears to be a difficult problem,
		it might be worth first restricting
		to the case of understanding the ``weak temporal tree-child 
		hybrid number"; how much different can 
		this number be from the rigid hybrid number, and can we decide when a 
		set of trees is weakly displayed by a temporal tree-child network?
		To answer these questions it could be worth first  trying to decide 
	whether or not two phylogenetic trees are rigidly displayed by some
	temporal tree-child network if and only if they are weakly displayed
	by some temporal tree-child network.\\
	
	\noindent {\bf Acknowledgement}
	KTH and VM thank the London Mathematical Society 
	and SL thanks the New Zealand Marsden Fund for their financial support.
	All authors thank the Biomathematics Research Centre, University of Canterbury, and 
	The Lorentz Center, Leiden, where they discussed parts of this work.

\bibliography{biblio}{}
\bibliographystyle{spbasic}      

\end{document}